\documentstyle[twocolumn,aps,epsf]{revtex}

\begin{document}
\draft

\newcommand{\mytitle}[1]{
\twocolumn[\hsize\textwidth\columnwidth\hsize
\csname@twocolumnfalse\endcsname #1 \vspace{1mm}]}

\mytitle{

\title{Fermi liquid to Luttinger liquid transition at the edge of a
two-dimensional electron gas}
\author{M. Hilke$^\flat$, D.C. Tsui, M. Grayson$^\ddag$,
L.N. Pfeiffer$^a$ and K.W. West$^a$}
\address{Dpt. of Elect. Eng., Princeton University, Princeton,
New Jersey, 08544}
\address{$^\ddag$Walter Schottky Institut, Technische
Universit\"at M\"unchen, Garching, 85748, Germany}
\address{$ ^{a}$Bell Laboratories, Lucent
Technologies, Murray Hill,  New Jersey, 07974}
\date{March 28, 2001}
\maketitle

\begin{abstract}
We present experimental results on the tunneling into the edge of
a two dimensional electron gas (2DEG) obtained with a GaAs/AlGaAs
cleaved edge overgrown structure in a strong perpendicular
magnetic field. While the 2DEG exhibits typical fractional quantum
Hall features of a very high mobility sample, we observe the onset
of a non-linear current-voltage characteristic in the vicinity of
$\nu=1$. For filling factor $\nu<1$ the system is consistent with
a non-Fermi liquid behavior, such as a Luttinger liquid, whereas
for $\nu>1$ we observe an Ohmic tunneling resistance between the
edge and a three dimensional contact, typical for a Fermi liquid.
Hence, at the edge, there is a transition from a Luttinger liquid
to a Fermi liquid. Finally, we show that the Luttinger liquid
exponent at a given filling factor is not universal but depends on
sample parameters.

\end{abstract}

}

In one dimension and in the presence of interactions, a metal can
have a Fermi surface in agreement with Luttinger's theorem
\cite{luttinger}. However, fermionic quasi-particles are no longer
possible and the elementary excitations are replaced by bosonic
charge and spin fluctuations dispersing with different velocities.
Hence, this one-dimensional metal is no longer a Fermi-liquid but
a Luttinger liquid (LL) \cite{haldane}. Models describing
one-dimensional interacting Fermions were first considered by
Tomonaga and Luttinger \cite{tomonaga}.

In a pioneering work on the fractional quantum Hall (FQH) states,
Wen \cite{wen} showed that the edge modes can be represented as
chiral $LL$s. The chirality is due to the presence of a magnetic
field, which forces the edge states to propagate in one direction.
A unique feature of the {\em chiral} $LL$ is the absence of
back-scattering, i.e., no localization can occur. A key
theoretical result is the existence of power-law correlation
functions, which lead to the vanishing of the momentum
distribution function at $k_F$ following a power-law, i.e.,
$n(k)\sim|k-k_F|^\alpha$, where $\alpha$ is related to the
interaction strength. As a consequence, the tunneling
current-voltage (I-V) characteristics follows $I\sim V^\alpha$
\cite{wen}. For the particular case, where the filling factor
$\nu=1/3$, Wen predicted that $\alpha=3$, hence the tunneling
current should vanish like $I\sim V^3$ when tunneling from a Fermi
liquid into a Luttinger liquid. This is very different from the
Fermi liquid-to-Fermi liquid tunneling which would be Ohmic.

Following the predictions of Wen \cite{wen} and others
\cite{kane}, several experimental attempts were made in order to
observe this power-law dependence. The first experiments
considered a gate induced constriction to tunnel between two FQHE
liquids \cite{proexp,conexp}. Unfortunately, in some cases the
results were consistent with a power-law \cite{proexp} but not in
others \cite{conexp}. This was largely attributed to the
smoothness of the potential barrier causing the possible
reconstruction of the edge and an energy dependent tunneling
barrier. Chang et al. \cite{chang} avoided this problem by growing
a sharp tunneling barrier on the cleaved edge of a two dimensional
electron gas (2DEG). They obtained a good power-law over more than
a decade in voltage to obtain a tunneling exponent ($\alpha\simeq
2.7$ at $\nu=1/3$) close to Wen's prediction.

When moving away from the primary fraction $\nu=1/3$ to
$\nu=p/(2np\pm 1)$ (where p and n are positive integers), the edge
cannot be described anymore by a single LL edge mode but requires
several additional modes, the number and nature of which depends
strongly on the particular fraction and, moreover, the disorder
becomes important because of possible inter-mode scattering. The
overall structure of these states is reviewed in ref. \cite{wen2}.
As a consequence, the recent experimental result from Grayson et
al. \cite{grayson} came as a surprise, because instead of
observing a plateau-like structure between $\nu^{-1}=2$ and 3, as
expected from both the composite fermion theory
\cite{shytov,levitov} and a disordered edge in the hierarchical
model \cite{polchinski}, they observed a linear dependence of the
exponent on the inverse filling factor, $\alpha\simeq\nu^{-1}$.
Recent theories have attempted to account for this behavior using
different approaches \cite{theopro} and are currently under
debate. Subsequent experimental work, indicated a weak plateau
feature at $\alpha\simeq 2.7$, suggesting a stable single edge
mode, but for $\nu^{-1}=3$ or 4.5, depending on the sample
\cite{chang2}. At low filling factors, a tunneling resonance
consistent with a Luttinger liquid behavior was observed recently
\cite{matt2}.

In this article, we probe for the first time the edge of the 2DEG
over a large range in filling factors from $\nu=1/3$ to $B=0$,
hence also higher Landau levels. In order to achieve this we start
with a very high mobility 2DEG ($\mu>10\times 10^6$cm$^{-2}$/Vs)
confined in a symmetrically doped GaAs/AlGaAs quantum well. The
quantum well is then placed in the molecular beam epitaxy (MBE)
chamber and cleaved for a subsequent growth along the (110)
direction. First an atomically sharp barrier of
(Al$_{x}$Ga$_{1-x}$As) is grown and then a 5000 \AA $ $ n-doped
GaAs layer \cite{loren}. In order to probe a large voltage range,
we fixed the height of the barrier at a high value, i.e, 200meV by
using $x=20\%$ as barrier material. The low temperature (30mK) and
zero field tunneling resistance can be varied by using different
barrier widths (20-120\AA). For the thinnest barrier (20 \AA $ $)
the tunneling resistance is even smaller than the 2D resistance,
whereas for the thickest 120 \AA $ $ barrier the tunneling
resistance is 200k$\Omega$. Therefore, most of the results in this
work are based on the 60 \AA $ $ and 120 \AA $ $ barriers as they
cover our $B$-field range of interest, including the low field
regime. Thanks to the high tunneling resistance we were able for
the first time to measure the 3D Shubnikov-de Haas oscillations
directly on the edge because the 2D does not short the edge. This
enables us to extract the effective 3D Fermi energy directly. In
Fig. 1, we have plotted the magnetoresistance of the edge,
measured with an AC resistance bridge and obtain a 3D Fermi energy
of 33 meV ($n_{3D}=4.5\times 10^{17} cm^{-3}$). This can be
compared to the Fermi energy of the 2DEG, which is 13 meV for a
2DEG density of n=$2.2\times 10^{11}$cm$^{-2}$ in a 300 \AA $ $
quantum well (6 meV). Hence, the Fermi energy mismatch is 20 meV.
The value of the mismatch is important when evaluating the
internal electrical field building up at the interface due to the
mobile carriers. This bend bending could affect the density at the
edge of the 2DEG compared to the density of the bulk 2DEG

\vspace*{-.3cm}
\begin{figure}
\epsfsize=0.65 \columnwidth
\epsfbox{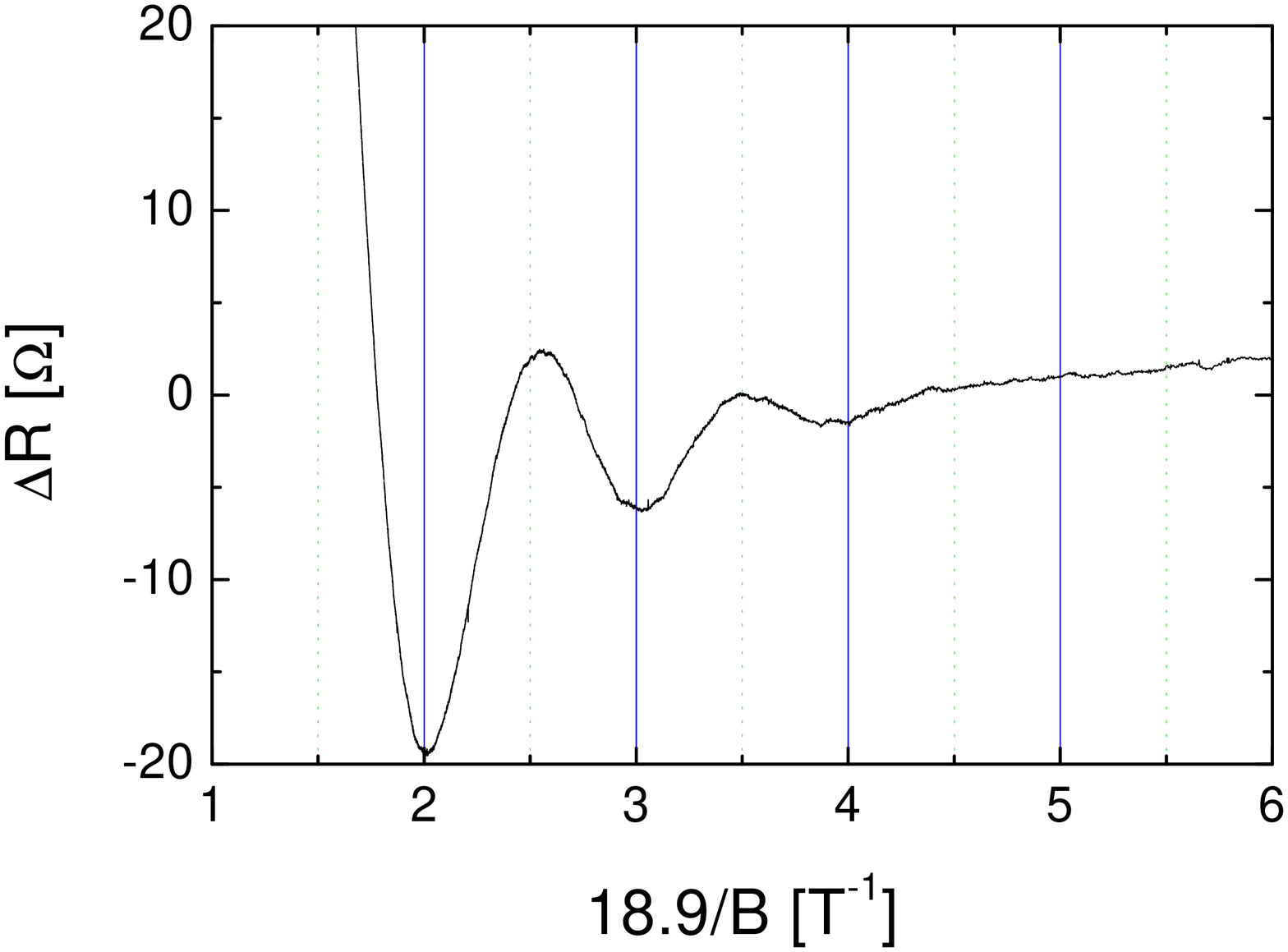}
\caption{
Magnetoresistance measured directly on the edge as a function of
the normalized inverse magnetic field at 30mK. The contacts used
were 1 and 8 as represented in the sketch of fig. 2.}
\end{figure}

The magnetoresistance trace of the 2DEG system is shown in Fig. 2
and exhibits typical features between Landau levels characteristic
of a very high mobility 2DEG. In the same figure, the tunneling
resistance across the barrier, between the 2DEG and the edge, is
plotted as a function of the magnetic field for different
voltages. The overall magnetic field dependence is dominated by an
exponential increase of the resistance as a function of $B$. Two
effects contribute to this increase in resistance. Indeed, for a
perfect barrier the momentum conservation parallel to the barrier
is suppressed by a quantizing magnetic field. However, scattering
along the barrier can reduce this effect. Second, the penetration
length orthogonal to the barrier is exponentially suppressed by
the square of the magnetic length.

The most important feature in this figure is that below 6.5 T the
tunneling resistance is independent on the voltage, whereas above
6.5 T the resistance is strongly voltage dependent, i.e., the
resistance is not Ohmic. When using a narrower barrier, such as 60
\AA, we obtain the same $B$-field for the onset of non-linearity,
but at a much smaller tunneling resistance. Hence, the onset of
non-linearity only depends on $B$ and not on the value of the
tunneling resistance. This overall behavior is very suggestive to
a transition from a Fermi liquid (Ohmicity) to a LL, where
$B_C\simeq6.5$ T would be the critical field of the transition.
This $B$-field corresponds to a filling factor of $\nu=1.3$. In
order to investigate this behaviour further we now analyze the
non-linearity in more detail and compare it with Wen's \cite{wen}
theoretical prediction for a chiral LL.

\vspace*{-.3cm}
\begin{figure}
\epsfsize=1 \columnwidth
\epsfbox{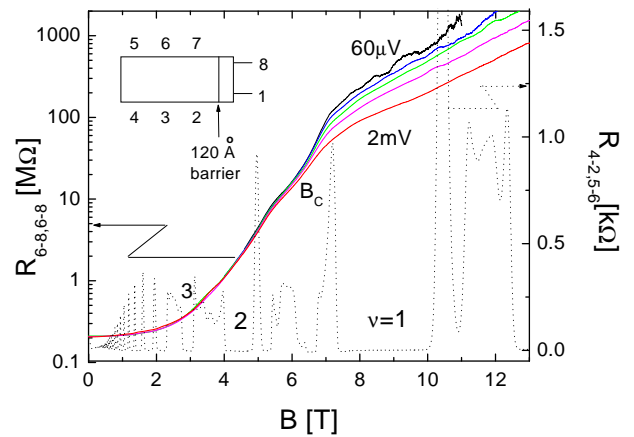}
\vspace*{.1cm}
\caption{ The magnetoresistance, $R_{4-2,5-6}$, of the 2DEG as a
function of the magnetic field is plotted in dotted lines. The
tunneling resistance, $R_{6-8,6-8}$, is plotted for voltages of
2,1,0.5,0.16,0.06 mV applied across the barrier. $R_{4-2,5-6}$ was
obtained by measuring the voltage between contacts 5 and 6 and
applying an AC ($<5Hz$) current ($10nA$) between contacts 4 and 2.
For $R_{6-8,6-8}$ we applied a negative DC bias on contact 8
corresponding to the edge. All traces were obtained at 30 mK. }
\end{figure}

We start by plotting in Fig. 3 the I-V characteristics across the
barrier for different $B$'s at 30 mK. The I-V's are obtained by
measuring the DC traces, unlike refs. \cite{chang,grayson}, where
an AC voltage was applied. For clarity we have only plotted the IV
traces, where a positive bias was applied to the edge, i.e.,
electrons tunnel out of the 2DEG. The negative bias data is
identical up to a voltage of about 2 mV, above which strong
asymmetries arise. For a more detailed discussion of these
asymmetries beyond 2 mV the reader is referred to \cite{hilke}.
When limiting our range of interest from 0.2 to 2 mV, two regimes
can be identified: for $B\leq B_C$, where $B_C\simeq 6.5$ T, the
I-V's are linear, but for $B>B_C$ the I-V's follow a power-law
larger than 1. To extract the power-law exponent, $\alpha$, we
have performed a least square fit including all data in that
range. The fits are shown in dotted lines along with the data in
solid lines. The quality of the fit is very good over this voltage
range, but below 0.2 mV there are deviations from the power-law,
which can be attributed to finite temperature effects and to
limitations in our experimental sensitivity. Above 2 mV we also
have deviations from a power-law, which are probably not related
to the LL behavior, because similar deviations are also seen at
B=0 and are not symmetric with respect to the sign of the bias.
For the data above 6.5 T the shapes of the IV's are very similar
to the ones in refs. \cite{chang,grayson}, which were attributed
to the LL behavior.

When further increasing $B$, the power-law exponent $\alpha$
increases gradually and is plotted in Fig. 4 for two different
barriers 120 \AA $ $ and 60 \AA. Below 6.5 T, $\alpha$ is
essentially constant. The crossing of the linear extrapolation
between the points above 7 T (dotted line) and $\alpha=1$ is at
6.7 T, which is very close to the onset of nonlinearity $B_C=6.5$
T of Fig. 2. Interestingly, the small difference in resistance at
different voltages between 4.5 and 6.5 T in Fig. 2 is reflected in
Fig. 4, by a value of $\alpha$ slightly above 1, for the same
$B$-field range.

\vspace*{-.3cm}
\begin{figure}
\epsfsize=1 \columnwidth
\epsfbox{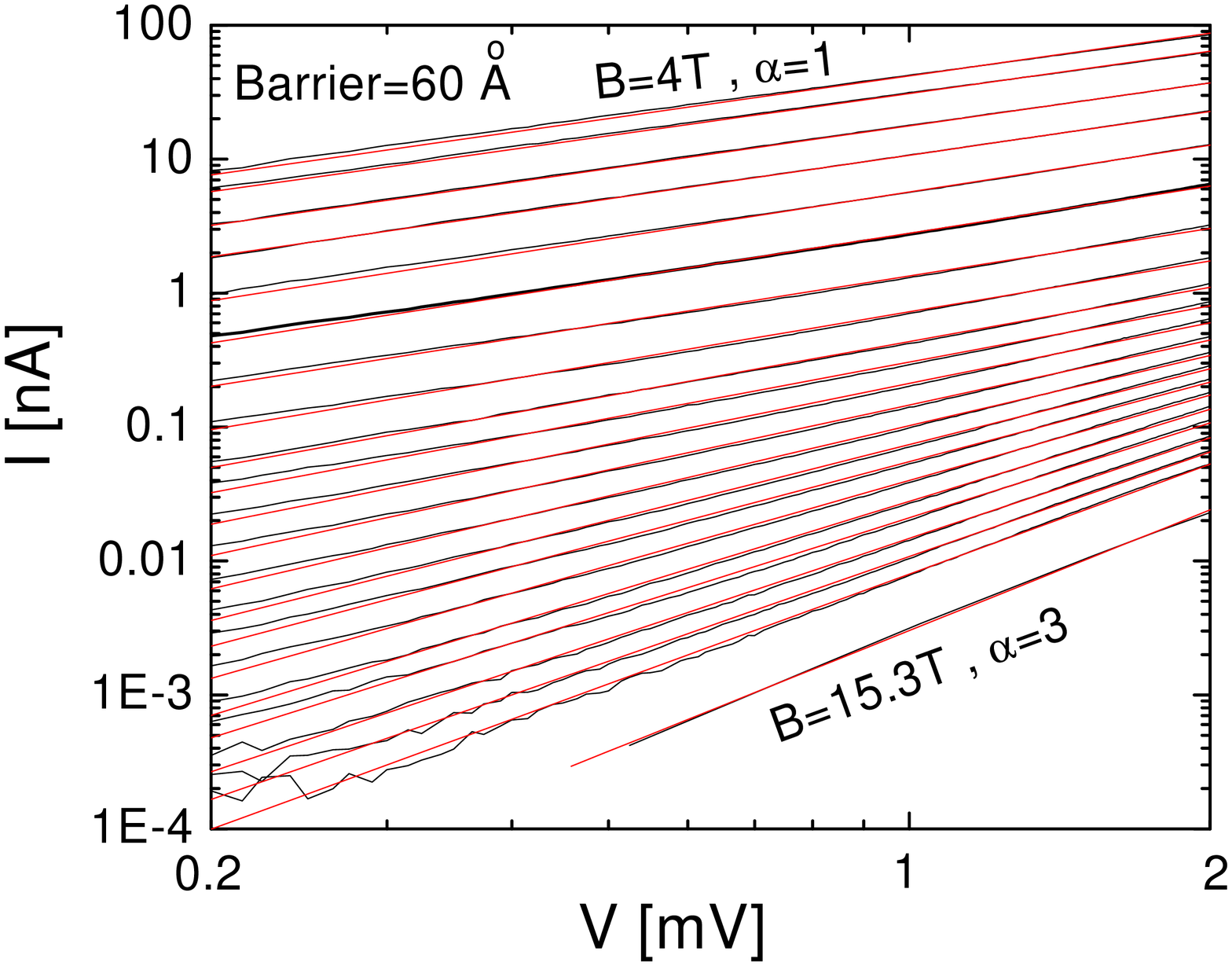}
\caption{ Current
voltage characteristics on a logarithmic scale at B=4, 4.5, 5,
..., 14, 15.3 T. The thick line corresponds to B=6.5 T. The dotted
lines are the linear log-log fits to the data.}
\end{figure}

For reference, we have also plotted the results for the 120 \AA $
$ barrier width in Fig. 4. Although, the two barriers (60 and 120
\AA ) have very different tunneling resistances, the overall
dependence of $\alpha$ is very similar. Because the tunneling
resistance of the 60 \AA $ $ barrier is much smaller we can
extract the exponent $\alpha$ to a larger $B$. The experimental
limit is given by our current noise level (10-100fA). For fields
below 4 T the tunneling resistance of the 60 \AA $ $ barrier
becomes comparable to the two-terminal resistance of the 2D,
implying that the barrier resistance can no longer be extracted.
In contrast, the 120 \AA $ $ barrier has a tunneling resistance
much larger than the two-terminal resistance even below 4 T (at
B=0 the tunneling resistance is still 200k$\Omega$). Therefore,
using the 120 \AA $ $ barrier we can also cover all the low field
behavior. The error bars shown in Fig. 4 are given by the mean
square deviation from a power-law. The extraction of the exponent
is very robust when comparing positive to negative bias. Indeed,
when extracting the exponents from the negative bias IV's (not
shown) the exponents fall within less than 10 \%.

\begin{figure}
\epsfsize=1 \columnwidth
\epsfbox{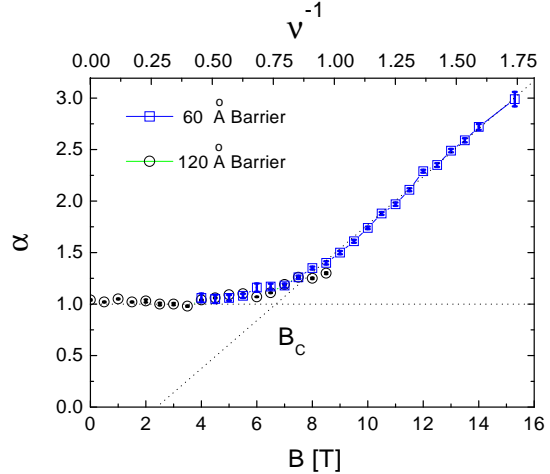}
\caption{Exponent
$\alpha$ extracted from the power-law part of the I-V curves as a
function of $B$ and filling factor $\nu$. The circular dots are
for a 120\AA $ $ wide barrier and the square dots for a 60\AA $ $
wide barrier. }
\end{figure}

Thus, for $B>B_C$ the I-V's follow a power-law indicative of a
Luttinger liquid, whereas below $B_C$ we have a standard
Fermi-liquid, hence $B_C$ represents the transition between a
Fermi-liquid and a Luttinger-liquid. Moreover, this demonstrates
that below $B_C$ the rich structure of the fractional quantum Hall
system is dominated at the edge by a standard Fermi-liquid mode.
 In our samples this transition occurs at a filling factor of
$\nu_C\simeq 1.3$. This is very different to the value obtained in
refs. \cite{chang,grayson}, where the extrapolation to $\alpha=1$
would yield $\nu\simeq 0.73$. Further, we obtain in our case a
slope of $\alpha\simeq 2.0 \nu^{-1}-0.55$ whereas in ref.
\cite{grayson} they obtained $\alpha\simeq 1.16\nu^{-1}-0.58$.
Hence, experimentally, for a fixed 2DEG filling factor the value
of the exponent is dependent on sample parameters, although the
onset of non-linearity does not depend on the barrier width nor on
the magnitude of the tunneling resistance, when all other
parameters are the same (like for our samples with different
barrier widths 60\AA $ $ and 120\AA $ $). Hence, the value of the
exponent for a given filling factor is different to the chiral LL
predictions by Wen \cite{wen}.

A possible explanation could be a shift in the local density
distribution close to the edge. Indeed, our tunneling edge is at
the interface between two differently doped semiconductors, which
leads to the band bending within the characteristic depletion
length \cite{matt3}. In the simplest case of 2 adjacent 3D n-doped
semiconductors, the local electron density of the side with the
lowest Fermi energy is enhanced at the interface. For a 2D-3D
interface this is not necessarily the case anymore. Indeed,
Levitov and co-workers \cite{levitov} used a Thomas-Fermi model,
but without including quantum confinement effects, nor a magnetic
field, to calculate the density distribution of the 2DEG close to
the edge for a similar structure. They found that the 2D edge
density could be about $25\%$ smaller than in the bulk, when the
3D has a 20 meV higher Fermi energy than the 2D like in our case.
They further calculated that the 2D edge density increases
continuously with increasing 3D density. However, when comparing
two samples with the same 2D Fermi energy of 17 meV but two
different 3D Fermi energies (33 and 62 meV), we found a very
similar onset of non-linearity for both samples at $\nu\simeq
0.65$. This suggests that a more involved theoretical treatment is
necessary in order to account for a possible edge density
redistribution. In particular, it may be crucial to include the
effects of a large $B$-field as it is possible that the average
density at the edge is dependent on $B$.

Assuming, however, that we do have a different 2D filling factor
at the edge than in the bulk and that the effective $\nu_{edge}=1$
for the edge occurs at 6.7 T because the $\nu=1$ state is expected
to behave like a standard Fermi-liquid \cite{wen2}, we would
obtain an edge density $24\%$ smaller than the bulk density. For
the samples used by Grayson {\em et al.} \cite{grayson}, the
intercept where $\alpha=1$ would occur at $\nu^{-1}_{edge}=1$ if
the density at the edge of the 2DEG is assumed to be $40\%$ larger
than in the bulk. Rescaling our data by -24\%
($\nu_{edge}=0.76\times\nu$) and Grayson's data by +40\% we find
in both cases that $(\alpha-1)\simeq 1.6(\nu^{-1}_{edge}-1)$. This
is very intriguing, but it is not clear whether this is generic or
not. This shift in edge vs. bulk 2D density, could also explain
the behavior in ref. \cite{chang2}, where they observe a weak
plateau behavior for lower filling factors than expected.

In the following we compare our results with existing theories.
Most theoretical results fall in two main classes. In one group,
calculations are consistent with $(\alpha-1)=2(\nu^{-1}-1)$
\cite{wen,kane,wen2,shytov,levitov,polchinski} and in the other
group the calculations are consistent with
$(\alpha-1)=(\nu^{-1}-1)$ \cite{theopro}. However, none of the
experimental curves fall clearly on one of these theoretical
dependencies, even if we assume a shift in the edge versus bulk 2D
density. We believe that a more detailed analysis of the edge
distribution is needed in order to resolve this issue.

In conclusion, we have observed a Luttinger liquid - to - Fermi
liquid transition by tunneling into the edge of a 2DEG system. For
high filling factors, our results indicate that a Fermi-liquid
outer edge mode dominates the edge physics at higher filling
factors. We further showed that the Luttinger liquid exponent is
not universal in relation to the bulk two-dimensional filling
factor but that it can depend on other sample parameters. This
dependence, however, could be due to a shift in the 2D edge
density.

We would like to acknowledge Albert Chang, Leonya Levitov and
Shivaji Sondhi, for helpful discussions. This work was supported
in part by the National Science Foundation.

$^\flat$Permanent address: Dpt. of Physics, McGill University,
Montr\'eal, Canada H3A 2T8. E-mail: Hilke@physics.mcgill.ca

\end{document}